\documentclass[a4paper,12pt]{article}
\usepackage{graphicx}
\usepackage{amssymb,amsmath}
\usepackage{textcomp}
\usepackage{subfigure}
\usepackage{feynmp}
\usepackage{cite}
\def\pacs#1{\vspace{10pt} \hspace{0.33cm} \rm PACS numbers: #1 \par \vspace{10pt}}

\title{Fractal structure and non extensive statistics}
\author{Airton Deppman$^{1}$, Eugenio Meg\'ias$^{2,3}$, Debora P. Menezes$^{4}$, Tobias Frederico$^{5}$}
\date{1- Instituto de F\'isica -  Universidade de S\~ao Paulo \\ email: deppman@if.usp.br \\ Rua do Mat\~ao Travessa R Nr.187 CEP 05508-090 Cidade Universit\'aria, S\~ao Paulo - Brasil; \\ 2- Departamento de F\'{\i}sica Te\'orica, Universidad del Pa\'{\i}s Vasco UPV/EHU, Apartado 644, 48080 Bilbao, Spain;\\ 3- Departamento de F\'{\i}sica At\'omica, Molecular y Nuclear and Instituto Carlos I de F\'{\i}sica Te\'orica y Computacional, Universidad de Granada, Avenida de Fuente Nueva s/n,  18071 Granada, Spain. \\ 4- Departamento de F\'isica, CFM, Universidade Federal de Santa Catarina, CP 476, CEP 88040-900, Florian\'opolis, Brazil.; \\ 5- Instituto Tecnol\'ogico da Aeron\'autica.}

\begin{document}

\maketitle

\begin{abstract}
The role played by non extensive thermodynamics in physical systems has been under intense debate for the last decades. With many applications in several areas, the Tsallis statistics has been discussed in details in many works and triggered an interesting discussion on the most deep meaning of entropy and its role in complex systems. Some possible mechanisms that could give rise to non extensive statistics have been formulated along the last several years, in particular a fractal structure in thermodynamics functions was recently proposed as a possible origin for non extensive statistics in physical systems. In the present work we investigate the properties of such fractal thermodynamical system and propose a diagrammatic method for calculations of relevant quantities related to such system. It is shown that a system with the fractal structure described here presents temperature fluctuation following an Euler Gamma Function, in accordance with previous works that evidenced the connections between those fluctuations and Tsallis statistics. Finally, the fractal scale invariance is discussed in terms of the Callan-Symanzik Equation.
\end{abstract}


\section{Introduction}

As the formulation of new mathematical tools opens opportunities to describe systems of increasing complexity, entropy emerges as an important quantity in different areas. In recent years our knowledge about the role played by entropy in Physics as well as in other fields has increased rapidly in part, at least, due to the formulation of new entropic forms that generalize in some way the one first proposed by Boltzmann. The non additive entropy, $S_q$, introduced by Tsallis~\cite{Tsallis1988} has found wide applicability, triggering interesting studies on the most deep meaning of entropy and on its importance in the description of complex systems~\cite{BeckCohen,Thurner,HanelThurnerGellMann, Tempesta, Kalogeropoulos, Kalogeropoulos2}.

The full understanding of the non extensive statistics formulated by Tsallis, however, has not been accomplished as yet. Four different connections between Boltzmann and Tsallis statistics have been proposed so far~\cite{Borland, BeckTFluct, WilkWlodarczykTFluct, Biro, Deppman2016,Deppman:2016epj}, all of them giving a clear meaning to the entropic index, $q$, that appears in the non extensive case, and in all connections Boltzmann statistics is obtained as a special case. But it seems that the physical meaning of this parameter is not understood in the general case, and the difficulty to grasp the significance of the entropic index may be related to the fact that this quantity never appeared before in Thermodynamics while temperature, even if it appears as another parameter in Statistical Mechanics, had already an intuitive meaning in the description of thermodynamical systems. This fact, however, cannot diminish the importance of the index $q$ in the formulation and description of systems where Boltzmann statistics is not suitable.

In the present work we make a detailed analysis of the fourth of those connections, where a system featuring fractal distributions in its thermodynamics properties, that was named thermofractals\cite{Deppman2016}, has been shown to follow Tsallis statistics. These fractals are relatively simple systems: they are conceived as objects with an internal structure that can be considered as an ideal gas of a specific number of subsystems that are also fractals of the same kind. The self-similarity between fractals at different levels of the internal structure follows from its definition and reveals the typical scale invariance. It has been shown that thermodynamical systems with the structure studied in the present work show fractional dimensions~\cite{Deppman2016}, another feature shared with fractals in general. The fractal dimension can be related to the fact that the system energy is proportional to a power of the number of particles, this power being different from unit. This and other aspects of those systems will be discussed in the present work. 

Although the motivation that prompted the formulation of thermofractals was related to applications of Tsallis distributions in High Energy Physics~\cite{Bediaga,Beck,Sena,Deppman2012,Cleymans_Worku,AzmiCleymans2,DeBhaskar,Wong,WongWilk,WilkWlodarczyk,Lucas,Lucas2,WWselfsymmetry,Wilk2007,Wilk2009,Megias,Biro2017}, Hadron Physics~\cite{Megias,Deppman2014,Debora,Megias:2014tha,Deppman:2015cda, Grygorian}, Astrophysics~\cite{Debora,Megias:2014tha,Deppman:2015cda,PedroCardoso} and Cosmic Ray spectrum~\cite{Beck2018}, the system is in fact general and in principle could find applications in other fields. Being a way to relate formally Tsallis and Boltzmann statistics, the analysis of these fractals may shed some light on the open questions around the meaning of the entropic parameter and on the fundamental basis of the non extensive statistics, and in this way it can also contribute to a better understanding of entropy. In this regard, it is worth mentioning that fractals was one of the starting points for the formulation of the generalized statistics~\cite{Tsallis1988}. In spite our objective here being the study of the general properties of thermodynamical fractals, the results obtained in the present work offer a new perspective in the analysis of hadron structure. This new perspective will be exploited in a future paper.

This work is organized as follows: in section 2 the main aspects of thermodynamical fractals are reviewed; in section 3 the fractal structure is analyzed in detail; in section 4 a diagrammatic scheme that can facilitate calculations is introduced, and some examples are given; in section 5 it is shown that the temperature of the fractal system addressed here fluctuates according to the Euler Gamma Function, a kind of temperature fluctuation already associated to Tsallis statistics; in section 6 we analyse the system scale invariance in terms of the Calla-Symanzik equation, a result that may be of importance for application in Hadron Physics; in section 5 our conclusions are presented.

\section{Fractals and Tsallis Statistics}

From a mathematical point of view, the basic difference between Boltzmann and Tsallis statistics is the probability factor, $P(E)$, which is an exponential function of energy in the case of Boltzmann statistics, and in the non extensive statistics proposed by Tsallis is a function called q-exponential, given by 
\begin{equation}
 P(\varepsilon)= A \bigg[1+ (q-1) \frac{\varepsilon}{k\tau}\bigg]^{-1/(q-1)}\, \label{qexp}\,,
\end{equation}
where $T$ is the temperature, $k$ is the Boltzmann constant, $A$ is a normalization constant and $q$ is the so-called entropic factor, which is a measure of the deviation of the system thermodynamical behavior from the one predicted by the extensive statistics. 

The emergence of the non extensive behavior has been attributed to different causes: long-range interaction, correlations, memory effects,which would lead to a special class of Fokker-Planck equation that would lead to a non extensive behavior~\cite{Borland}, temperature fluctuation~\cite{BeckTFluct, WilkWlodarczykTFluct}, finite size of the system~\cite{Biro}. In this work we analyze in detail a thermodynamical system recently proposed that presents a fractal structure in its thermodynamical functions, what leads to a natural description of its properties in terms of Tsallis statistics~\cite{Deppman2016}, and we show that  such system presents a fractal structure in its momentum space.

In the study of the thermodynamical properties of the fractal system of interest here, an important quantity is
\begin{equation}
  \Omega=\int_0^{\infty} P(U) dU\,, \label{eqomega0}
\end{equation}
with $U$ being the system total energy, and $P(U)$ its probability density. The main characteristic of such system~\cite{Deppman2016} is that $\Omega$, which can be written in Boltzmann statistics as
\begin{equation}
 \Omega=\int_0^{\infty} \rho(U) exp\left(-\frac{U}{kT}\right) dU\,, \label{eqomega1}
\end{equation}
where $\rho(U)$ is a particular density of states characteristic of such fractal, results to be equivalent to the integration over all possible energies of the q-exponential function, that is,
\begin{equation}
  \Omega=\int_0^{\infty} A \left[1-(q-1) \frac{\varepsilon}{k \tau} \right]^{-1/(q-1)} d\varepsilon \,. \label{eqomegaqexp}
\end{equation}
This result shows, therefore, that for systems with a particular density of states that will be presented in the following, it is possible to show that Tsallis statistics can substitute Boltzmann statistics while all the details of the internal structure of the system is ignored. In particular, this system presents a fractal structure in some thermodynamical quantities, and consequently it shows an internal structure with self-similarity, i.e., the internal components are identical to the main system after rescaling.

The importance of this result is two-fold: in one hand it allows to understand the emergence of non extensivity and the applicability of Tsallis entropy becomes clear, with the entropic index, $q$, being given by quantities well defined in the Boltzmann statistics; on the other hand, the structure obtained resembles in many ways strongly interacting systems, where Tsallis statistics has been used, indeed, to describe experimental distributions~\cite{WWselfsymmetry,Tokarev,Zborovsky,Zborovsky2018}.

The particular fractal structure that leads to Tsallis statistics has a density of states given by
\begin{equation}
  \rho(\varepsilon,F)=  A^\prime F^{\frac{3N'}{2}-1}   \left[\tilde{P}(\varepsilon)\right]^{\nu} \,, \label{densityofstates}
\end{equation}
where $F$ and $\varepsilon$ are independent quantities and $A^\prime = A kT$. The remaining part of the total energy, $E=U-F$, is such that
\begin{equation}
 \frac{\varepsilon}{kT}=\frac{E}{F}\,. 
\end{equation}
The exponent $\nu$ in Eq.~(\ref{densityofstates}) is a constant that will be related, in the following, to the entropic index, and the distribution $\tilde{P}(\varepsilon)$ of the variable $\varepsilon$. Notice that the phase space corresponding to a variation $dU$ is given, in terms of the new variable, by $dU=dF d\varepsilon$, since the two variables are idependent.

Substituting Eq.~(\ref{densityofstates}) in Eq.~(\ref{eqomega1}) it follows that
\begin{equation}
 \Omega= \int_0^{\infty} \int_0^{\infty} A F^{\frac{3N}{2}-1}   exp\left\{-\frac{\alpha F}{kT}\right\} dF \left[\tilde{P}(\varepsilon)\right]^{\nu} d\varepsilon \label{omegagas}
\end{equation}
with $N=N'+2/3$ and $\alpha=1+\varepsilon/kT$. Observe that now we have integrations on the independent variables $F$ and $\varepsilon$. It will be clear in the next section that the integration in $F$ is equivalent to an integration on the compound system momenta, and that the integration on $\varepsilon$ is related to an integration over the compounding systems mass.

It is straightforward to verify that $\Omega$ reduces to Eq.~(\ref{qexp}) if $\tilde{P}(\varepsilon)$ is itself a q-exponential. In fact, defining
\begin{equation}
  \tilde{P}(\varepsilon)=\left(1+\frac{\varepsilon}{NkT} \right)^{-\frac{3N \nu}{2(1-\nu)}}\,, \label{preqexp}
\end{equation}
substituting Eq.~(\ref{densityofstates}) into Eq.~(\ref{eqomega1}) and integrating the last equation in $F$, it will result in Eq.~(\ref{eqomegaqexp}) when the following substitutions are made:
\begin{equation}
\begin{cases}
 & q-1=\frac{2}{3N}(1-\nu) \\
 & T= \frac{\tau}{N (q-1)}   \label{qtau}
\end{cases}
\,.
\end{equation}
With these substitutions the density distribution results to be
\begin{equation}
 \tilde{P}(\varepsilon)=\left[1+(q-1)\frac{\varepsilon}{k\tau}\right]^{-\frac{1}{q-1}}\,. \label{Tsallisqexp}
\end{equation}
Comparing Eq~(\ref{eqomega0}) and Eq~(\ref{eqomegaqexp}) one can see that the energy distribution of the system, $P(U)$ is equal to the probability density $\tilde{P}(\varepsilon)$. Hence the energy distribution of the system follows the same distribution of the energy distribution of the compound system internal energy, i.e.,
\begin{equation}
 P(U) \sim \tilde{P}(\varepsilon)\,. \label{similarity}
\end{equation}

This result shows that some properties of the main system are found also in its compound systems, a self-similarity property that are present in system with fractal structure. In fact the system described by the density of states given by Eq.~(\ref{densityofstates}) is a fractal~\cite{Deppman2016}, and below its structure is discussed in details. Moreover, the distribution given by Eq.~(\ref{Tsallisqexp}) is the well known Tsallis distribution, hence we can conlude that using Tsallis statistics all complexity of the fractal system is taken into account in a rather simple way, since from the non extensive entropy associated to that statistics all thermodynamics properties can be derives by the usual relations of thermodynamics~\cite{Curado,Plastino}.

\section{Fractal Structure}

The results obtained in the last section show that the system with the density of states given by Eq.~(\ref{densityofstates}) presents self-similarity, allowing one to interpret it as a fractal system. In this section such structure will be analyzed, and it will be shown that such system is a fractal in the energy-momentum space. Notice that Eq~(\ref{omegagas}) can be written as
\begin{equation}
 \Omega= \int_0^{\infty} \int_0^{\infty} (AkT)  F^{\frac{3N'}{2}-1} exp\left(-\frac{F}{kT}\right) dF exp\left(-\frac{E}{kT}\right) dE\,. \label{fractalgas}
\end{equation}

The most evident aspect of a fractal structure is its scale invariance. For the system studied here it means not only that the self-consistency relation represented by Eq.~(\ref{similarity}) must be valid, but also that for the kinetic energy, $F$, the distributions must be the same at all levels of the fractal structure. From Eq.~(\ref{densityofstates}), it follows that the distribution for $F$ is
\begin{equation}
 \omega(F)=\int_0^{\infty} A^\prime F^{\frac{3N}{2}-1} \exp\bigg(-\frac{F}{kT}\bigg) dF\,,
\end{equation}
which represents a Maxwellian distribution of energy. Therefore the scale invariance of thermofractals will be accomplished with the requirement that the kinetic energy distribution and the internal energy distribution are invariant under a scale transformation, so
\begin{equation}
 \frac{F^{(0)}}{T^{(0)}}=\frac{F^{(n)}}{T^{(n)}}\,,
\end{equation}
and
\begin{equation}
\frac{\varepsilon}{kT}=\frac{E^{(n)}}{F^{(n)}} 
\end{equation}
remains constant, hence
\begin{equation}
 \frac{E^{(0)}}{T^{(0)}}=\frac{E^{(n)}}{T^{(n)}}\,.
\end{equation}
Here and in what follows we use upper index $(0)$ to refer to quantities for the initial level of the thermofractal structure, or main system, and upper index $(n)$ to refer to quantities for the $n$-th level of the structure. The energy of the initial thermofractal, or main system, is $E=E^{(0)}$, and the temperature of the internal structure to this level is $T=T^{(1)}$.

It is interesting to express the scaling properties in terms of the fractal dimension. The fractionary dimension is one of the distinguishing properties of fractals and express the fact that some quantities do not scale as one could naively expect from the topological dimension of the system. In the present case, as it was shown in Ref.~\cite{Deppman2016}, energy and particle multiplicity do not increase in the same way, a behavior that is different of that found in an extensive ideal gas. In fact, using the fractal dimension found for such system one gets~\cite{Deppman2016}
\begin{equation}
  \lambda_n=\frac{E^{(n)}}{E^{(0)}}=\left(\frac{1}{N}\right)^{\frac{n}{1-D}}\,,
\end{equation}
where
\begin{equation}
 D=1+\frac{\log N'}{\log R} 
\end{equation}
is the fractal dimension. Here $R$ is the ratio between the internal energy of a component system and that of its parent system, and is given in terms of the parameters $q$ and $N$ by
\begin{equation}
 R=\frac{(q-1)N/N'}{3-2q+(q-1)N'} \,.
\end{equation}
 
The internal energy distribution scales by a factor
\begin{equation}
 \lambda_n=\frac{E^{(n)}}{E^{(0)}}=\frac{T^{(n)}}{T^{(0)}}=\lambda^n\,, \label{scalingfactor}
\end{equation}
defining the quantity $\lambda = 1/N^{\frac{1}{1-D}}$,
and
\begin{equation}
 T^{(n)}=\left(\frac{1}{N}\right)^{\frac{n}{1-D}} T\,. \label{Trelation}
\end{equation}
 Therefore, fractals with different internal energies present energy distributions that are similar and scales with the internal energy of the systems, that is,
\begin{equation}
 P \left(E^{(n)} \right) dE^{(n)}=\lambda_n P \left(E^{(0)}\right) dE^{(0)}\,. \label{energyscaling}
\end{equation}
Remarkably, as all energies are rescaled, it also happens $\varepsilon$ to be rescaled, therefore one has
\begin{equation}
 \frac{\varepsilon}{k\tau}=\frac{\varepsilon^{(n)}}{k\tau^{(n)}}\,, 
\end{equation}
with $\tau^{(n)}$ determined by Eqs.~(\ref{qtau}) and~(\ref{Trelation}). Thus the argument of the $q$-exponential function in the probability distribution $P(\varepsilon) = A e_q(-\varepsilon/(k\tau))=A \tilde{P}(\varepsilon)$ does not change from one level to the other. This is in fact the essence of self-similarity, and $P(\varepsilon)$ is the self-similar distribution. Another interesting feature is that
\begin{equation}
 A^{(n)}d\varepsilon^{(n)}=A d\varepsilon \,.
\end{equation}

In what follows the structure of the system just described will be investigated in details. For the sake of clarity the symbols
\begin{equation}
 \tilde P(\varepsilon)=\frac{P(\varepsilon)}{A}=e_q(-\varepsilon/(k\tau)) 
\end{equation}
will be used. Note that $\tilde P(\varepsilon) A d\varepsilon$ is dimensionless. Due to property 2 of thermofractals one has at the level $n-1$ of the fractal structure
\begin{equation}
     A^{(n)}dE^{(n)}=[\tilde{P}(\varepsilon)]^{\nu} Ad\varepsilon \, \left(\frac{F^{(n)}}{kT^{(n)}}\right)  \,, \label{recursiveness}
\end{equation}
where $F^{(n)}$ is the total kinetic energy of the compound fractals and $E^{(n)}=F^{(n)} \varepsilon^{(n)}/kT^{(n)}$ is their total internal energy. The following normalized energies will be adopted:
\begin{equation}
 f^{(n)}= \frac{F^{(n)}}{kT^{(n)}}
\end{equation}
and
\begin{equation}
 \epsilon^{(n)}=\frac{E^{(n)}}{kT^{(n)}}\,.
\end{equation}
with $n=1,2, \dots$ corresponding to the level of the fractal structure. Note that the normalized energies are dimensionless and scale invariant.

Given a fractal with non extensive temperature $\tau$, the system energy, $\varepsilon^{(n)}$, fluctuates according to the distribution
\begin{equation}
    P(\varepsilon^{(n-1)})d\varepsilon^{(n-1)}=\left[1+(q-1)\frac{\varepsilon}{k\tau}\right]^{-\frac{1}{q-1}} A d\varepsilon\,, 
\end{equation}
and generalizing Eq.~(\ref{omegagas}) to any level $n-1$  one can write (see Eq.~(\ref{uf4}) in the Appendix).
\begin{equation}
 \Omega_n=\frac{A}{\Gamma(3N'/2)} \int_0^{\infty} \int_0^{\infty} \left(f^ {(n-1)}\right)^ {\frac{3N'}{2}-1} e^{-\alpha f^{(n-1)}} \left[\tilde P(\varepsilon)\right]^{\nu} f^{(n-1)} d\varepsilon  d f^{(n-1)} \,, \label{level1}
\end{equation}
$\Omega_n$ represents the energy distribution of a constituent fractal at the $n$-th level of the main system.

Let $f_i^{(n)}$ correspond to the kinetic energy of the $i$-th constituent fractal at the $n$-th level of the fractal structure, each one having an internal energy determined by $\epsilon^{(n)}=\varepsilon f^{(n)}_i$. Eq.~(\ref{level1}) can be written in terms of the kinetic and internal energy of each constituent fractal, since
\begin{equation}
\int_0^{\infty} \int_0^{\infty} \left[f^{(n-1)} \right]^{(3N'/2)-1} e^{-f^{(n-1)}} e^{-\epsilon^{(n-1)}} \left[\tilde P(\varepsilon)\right]^{\nu} A d\varepsilon \, f^{(n-1)} df^{(n-1)} \,.
 \label{nleveltransition}
\end{equation}
Note that 
\begin{equation}
 A d\varepsilon \, f^{(n-1)} = A d\varepsilon \frac{F^{(n-1)}}{kT^{(n-1)}}=A\lambda_n^{-1} \frac{d\varepsilon}{kT} F^{(n-1)}\,, \label{epsilonvarepsilon}
\end{equation}
therefore also the constant $A$ scales as
\begin{equation}
 A^{n}=A\lambda_n^{-1}\,, \label{Ascale}
\end{equation}
with $A^{0}=A$ being the constant for the main system. This result is consistent with the temperature scale in Eq.~(\ref{Trelation}) and with the energy scaling relation in Eq.~(\ref{energyscaling}). It results that
\begin{equation}
 A d\varepsilon f^{(n-1)}=A_{n-1} d\epsilon^{(n-1)}\,, 
\end{equation}
with $\epsilon^{(n-1)}$ the normalized total internal energy of the thermofractals at the level $n-1$. Of course
\begin{equation}
 \epsilon^{(n-1)}=\sum_{i=1}^{N'} \epsilon^{(n)} \,.
\end{equation}
The term $d\epsilon^{(n-1)}$ can be written in terms of $d\epsilon^{(n)}$ as
\begin{equation}
 d\epsilon^{(n-1)}=\left[ \prod_{i=1}^{N'} \int_0^\infty d\epsilon_i^{(n)} \delta \left(\epsilon^{(n-1)}-\sum_{j=1}^{N'}\epsilon_j^{(n)}\right) \right] d\epsilon^{(n-1)} \,, 
\end{equation}
since it is related to the number of possible states $\{\epsilon^{(n)}_i\}$ that would result in the total energy $\epsilon^{(n-1)}$.

With these definitions one has
{ \footnotesize
\begin{equation}
 \begin{split}
& P(\epsilon^{(n-1)},f^{(n-1)}) d\epsilon^{(n-1)} df^{(n-1)}  =
\left[f^{(n-1)} \right]^{(3N'/2)-1} e^{-f^{(n-1)}} e^{-\epsilon^{(n-1)}} \left[\tilde P(\varepsilon)\right]^{\nu} A d\varepsilon \, f^{(n-1)} df^{(n-1)} = \\
& \left[f^{(n-1)} \right]^{(3N'/2)-1} e^{-f^{(n-1)}} e^{-\epsilon^{(n-1)}}  A^{(n-1)}d\epsilon^{(n-1)} df^{(n-1)} = \\
& \frac{\Gamma(3N'/2)}{[\Gamma(3/2)]^{N'}} \left[ \prod_{i=1}^{N'} \label{nlevel}  
\int_0^\infty \left(A^{(n)}kT^{(n)}\right) d\epsilon_i^{(n)} \int_0^{\infty} 
df_i^{(n)}  \left[f_i^{(n)} \right]^{\frac{3}{2}-1} e^{-f_i^{(n)}} e^{-\epsilon_i^{(n)}} 
\delta_{f,n}  \delta_{\epsilon,n}  \right]  d\epsilon^{(n-1)} df^{(n-1)} 
\end{split} 
\end{equation}
}
where
\begin{equation}
 \delta_{\epsilon,n}=\delta\left(\epsilon^{(n-1)} - \sum_i \epsilon_i^{(n)}\right)
\end{equation}
and 
\begin{equation}
 \delta_{f,n}=\delta\left(f^{(n-1)}-\sum_i f_i^{(n)}\right)\,.
\end{equation}
Observe that the integrations inside brakets are performed on the variables corresponding to the level $n$.

In Eq.~(\ref{nlevel}) it was used relation~(\ref{epsilonvarepsilon}) for writing $d\epsilon_i$ in place of $d\varepsilon$.
Since
\begin{equation}
 f_i^{(n)}=\frac{[p_i^{(n)}]^2} {2 m_i^{(n)}kT^{(n)}}\,,
\end{equation}
with $m_i^{(n)}$ being the mass of the $i$-th coonstituent fractal. One can identify mass with internal energy of the fractal, so $m_i^{(n)}=\epsilon^{(n)}kT^{(n)}$, so it follows that 
\begin{equation}
 f_i^{(n)}=\frac{[\pi^{(n)}_i]^2}{2 \epsilon_i^{(n)}}\,,
\end{equation}
where
\begin{equation}
 \pi^{(n)}_i=[p_i^{(n)}/kT^{(n)}] \,.
\end{equation}

Then Eq.~(\ref{nlevel}) results in (see Appendix, Eqs.~(\ref{uf1}) and~(\ref{uf4}))
\begin{equation}
P(f^{(n-1)},\epsilon^{(n-1)}) = \prod_{i=1}^{N'}  \left(A^{(n)}kT^{(n)}\right) \int_0^{\infty} d\epsilon_i^{(n)} \int_{-\infty}^{\infty} d^3 \pi_i^{(n)} (2\pi \epsilon_i^{(n)})^{-3/2} e^{-u_i^{(n)}}  \delta_{f,n} \delta_{\epsilon,n} 
\,, \label{1level}
\end{equation}
where $u_i^{(n)}=f_i^{(n)}+\epsilon_i^{(n)}$. In Eq.~(\ref{1level}) the potential $\Omega$ is described entirely in terms of the characteristics of the $N'$ compound thermofractals at the $n$-th level of the fractal structure, with $f_i^{(n)}$ and $\epsilon_i^{(n)}$ being related to their kinetic and internal energies, respectively.
But
\begin{equation}
 A^{(n)}kT^{(n)}=AkT=(2-q)/[N(q-1)] \,,
\end{equation}
$\epsilon_i^{(n)}=\epsilon_i$ and $\pi_i^{(n)}=\pi_i$ are independent of $n$, so it results
\begin{equation}
P(f^{(n-1)},\epsilon^{(n-1)}) = \prod_{i=1}^{N'} \frac{2-q}{N(q-1)}  \int_0^{\infty} d\epsilon_i \int_{-\infty}^{\infty} d^3 \pi_i (2\pi \epsilon_i)^{-3/2} e^{-u_i}  \delta_{f,n} \delta_{\epsilon,n} 
\,, \label{anylevel}
\end{equation}

The self-similar relation present in the fractal structure can be more apparent if Eq.~(\ref{anylevel}) is written as
\begin{equation}
P(f^{(n-1)},\epsilon^{(n-1)}) = \prod_{i=1}^{N'} \int_0^{\infty} A^{(n)} e^{-\epsilon_i^{(n)}} dE_i^{(n)} \int_{-\infty}^{\infty} d^3 \pi_i^{(n)} (2\pi \epsilon_i^{(n)})^{-3/2} e^{-f_i^{(n)}} \delta_{f,n} \delta_{\epsilon,n}
\,, \label{2level}
\end{equation}
where it is possible to recognize in the term $A^{(n)} dE_i^{(n)}$ the same expression as in Eq.~(\ref{recursiveness}), what allows the extension of calculations to include quantities of the next level in the fractal structure, i.e., level $n+1$, since 
\begin{equation}
 A^{(n)} dE_i^{(n)}=A^{(n+1)} dE_i^{(n+1)}
\end{equation}
 and
 \begin{equation}
 e^{-\epsilon_i^{(n)}}=e^{-\epsilon_i^{(n+1)}}\,.  
 \end{equation}
In addition, due to Eq.~(\ref{omegagas}),
\begin{equation}
 A^{(n+1)} e^{-\epsilon_i^{(n+1)}} dE_i^{(n+1)}=A^{(n+1)} \left(F^{(n+1)}\right)^{\frac{3N}{2}-1} \exp\bigg(-\frac{\alpha F^{(n+1)}}{kT^{(n+1)}}\bigg) dF^{(n+1)} [\tilde{P}(\varepsilon)]^\nu d\varepsilon^{(n+1)}\,, \label{ex1}
\end{equation}
what allows the passage to the next level by following all the steps described above. Before going into further calculations, however, a diagrammatic description will be introduced.

\section{Diagrammatic representation}

It is possible to have a diagrammatic representation of the probability densities that can facilitate calculations of
$\Omega$ and other relevant quantities. In Fig.~1 the
basic diagram symbols are presented, adopting $N'=2$ for simplicity. Each of the basic diagrams corresponds to a mathematical expression, and the correspondence can be established as follows:

\begin{figure}[ht]
 \centering
 \includegraphics[scale=1]{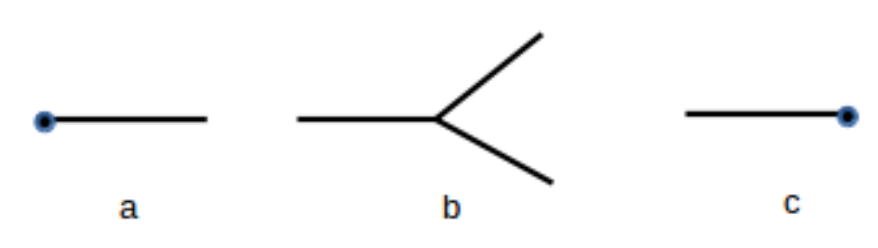}  \label{graphs1}
 \caption{Basic diagrams for the fractal structure: (a) main fractal; (b) vertex; (c) final fractal}
\end{figure}

\begin{enumerate}
 \item A line corresponds to a term 
 \begin{equation}
  \int_{-\infty}^{\infty} d^3\pi \epsilon^{-3/2} e^{-f}\,, 
 \end{equation}
with $f=\pi^2/(2\epsilon)$ and $\epsilon=(u-f)$, where $u$ is the total energy of the fractal represented by the line.
 \item A vertex corresponds to the term 
   \begin{equation}
   (2\pi)^{-3/2} \prod_{i=1}^{N'} \, \delta \left(f-\sum_{j=1}^{N'}f_j \right) \, 
   \end{equation}
 \item To each final line, i.e., those lines that do not finish in a
   vertex, the associated  term reads
   \begin{equation}
      \int_{0}^{\infty} A\,kT  \, e^{-\epsilon} \left[\tilde P(\varepsilon)\right]^{\nu} d\epsilon\,.
   \end{equation}
\end{enumerate}

The simplest diagram of interest is a line with a vertex where each branch is a final line. In this case the diagram scheme results in
{\footnotesize 
\begin{equation}
 \int_{-\infty}^{\infty} d^3\pi^{(n)} (\epsilon^{(n)})^{-3/2} e^{-f^{(n)}} (2\pi)^{-3/2} \prod_{i=1}^{N'} \, \delta_{f,n} \delta_{\epsilon,n} 
 \int_{-\infty}^{\infty} d^3\pi_i^{n+1} (\epsilon_i^{(n+1)})^{-3/2} e^{-f_i^{n+1}} \int_{0}^{\infty} A\,kT  \, e^{-\epsilon_i^{n+1}} \left[\tilde P(\varepsilon)\right]^{\nu} d\epsilon_i^{n+1}\,. 
\end{equation}
}
Delta functions can be included to fix energy and momentum of some of the fractals at any level.

\begin{figure}[!ht]
    \centering
    \includegraphics[scale=0.7]{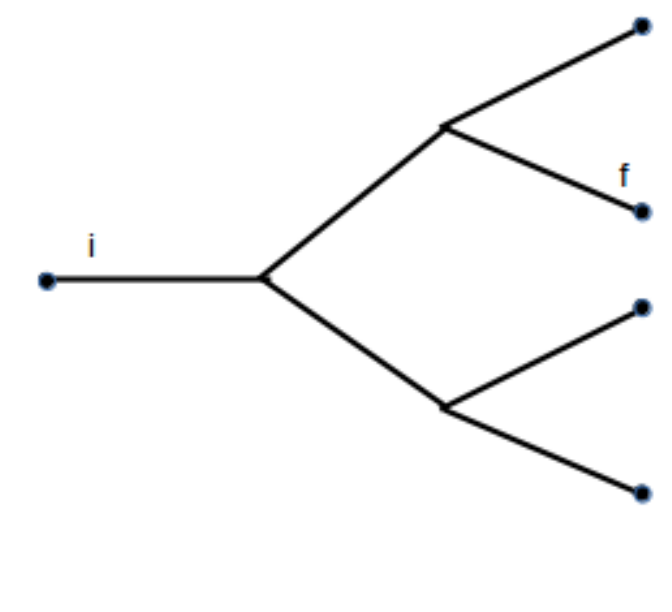}
    \caption{Example of a tree graph representing the different levels of a fractal.}
    \label{fig:ex2}
 \end{figure}

 As an example, consider the graph shown in
 Fig.~\ref{fig:ex2}. Observe that  there are two levels of the structure, the initial fractal has well defined
 momentum (it is indicated by $i$), and in the second level one of the
 systems has well defined energy and momentum. Such a diagram gives the probability to find a constituent fractal $f$ at the third level of the initial fractal $i$. According to the diagram rules one has
 {\footnotesize 
 \begin{equation}
  \begin{split}
 P_{i,f}= & \prod_{i=1}^{N'} \, \delta_{f_i,1} 
 \int_{-\infty}^{\infty} d^3\pi_i \epsilon_i^{-3/2} e^{-f_i}
 \prod_{j=1}^{N'} \, \delta_{f_{i,j},2} 
 \int_{-\infty}^{\infty} d^3\pi_{i,j} \epsilon_{i,j}^{-3/2} e^{-f_{i,j}}
 \int_{0}^{\infty} A\,kT  \, e^{-\epsilon_{i,j}} \left[\tilde P(\varepsilon)\right]^{\nu} d\epsilon_{i,j} \delta_{f_{1,2},f_f} 
  \end{split}
 \end{equation}
}
where $\delta_{f_{1,2},f_f}$ determines the kinetic part of the fractal indicated by $f$ at the second level.
 

It is possible to consider the fractal structure in the opposite way: given $N'$ fractals with energies $\{f_1,\epsilon_1, \dots\,f_{N'},\epsilon_{N'}\}$ varying in the range $df_1, f\epsilon_1, \dots, df_{N'},d\epsilon_{N'}$, the probability that they form a single fractal with energies $f=f_1+\dots+f_{N'}$ and $\epsilon=\epsilon_1+\dots+\epsilon_{N'}$ is given by
\begin{equation}
 P(E)dE=f^{(3N'/2)-1}e^{-f} e^{-\epsilon} \left[\tilde P(\varepsilon)\right]^{\nu} df A d\varepsilon\,, \label{merge1}
\end{equation}
with $E/kT=f+\epsilon$ and $\varepsilon=(\epsilon/f)kT$. This result
is a direct consequence of the fact that thermofractals are systems in
thermal equilibrium. After integrating on $f$ one obtains
\begin{equation}
   P(E)dE=P(\varepsilon)d\varepsilon\,, 
\end{equation}
showing the consistency of the fractal description introduced in the present work.

 
 \begin{figure}[!ht]
    \centering
    \includegraphics[scale=0.7]{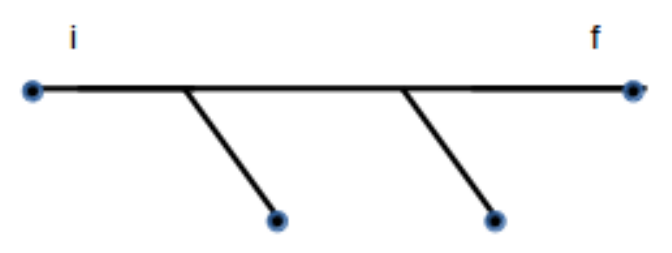}
    \caption{The same diagram of Fig. 2 represented as a linear graph. This is possible by rearranging terms in the summation of different contributions and using the merging property of thermofractals.}
    \label{fig:ex3}
 \end{figure}
 
The process described by Eqs.~(\ref{merge1}) corresponds to $N'$ fractals merging into a single one. In the example given above
and described in Fig.~\ref{fig:ex2}, the final systems generated from
the lower branch at the first level can be merged into a single
fractal. The tree diagram can then be reduced to a linear diagram, as shown in Fig.~\ref{fig:ex3}, resulting in a simpler expression for the probability calculated in that example. In this case the result is 
{\footnotesize
 \begin{equation}
  \begin{split}
 P_{i,f}= & \prod_{i=1}^{N'} \, \delta_{f_i,1} \delta_{\epsilon_i,1} \int_{-\infty}^{\infty} d^3\pi_i \epsilon_i^{-3/2} e^{-f_i}
 \prod_{j=1}^{N'} \, \delta_{f_{1,j},2} \delta_{\epsilon_{1,j},2} \int_{-\infty}^{\infty} d^3\pi_{1,j} \epsilon_{1,j}^{-3/2} e^{-f_{1,j}} \, \times \\ &
 \int_{0}^{\infty} A\,kT  \, e^{-\epsilon_{1,j}} \left[\tilde P(\varepsilon)\right]^{\nu} d\epsilon_{1,j}    \delta_{\epsilon_{1,2},\epsilon_f} \delta_{f_{1,2},f_f}
  \end{split}
 \end{equation}
}

\section{Temperature fluctuation in thermofractals}

In the right hand side of the last equality in Eq.~(\ref{nlevel}) the distribution of the kinetic energy of the thermofractals at the $n$th level is given by
\begin{equation}
 P(f_i^{(n)}) df_i^{(n)} =  \left[f_i^{(n)} \right]^{\frac{3}{2}-1} e^{-f_i^{(n)}}   \,, \label{Tfluct1}
\end{equation}
with 
\begin{equation}
 f_i^{(n)}=\frac{F_i^{(n)}}{T^{(n)}}
\end{equation}
where $T^{(n)}$ is the scaled temperature at the $n$th level of the thermofractal. However, at the level $n-1$ there are $N'$ thermofractals, and each of them present different internal energies. One could, therefore, write the temperature $T_j^{(n)}$ associated to  the thermofractal $j$ at the previous level. Then Eq.~(\ref{Tfluct1}) can be written as
\begin{equation}
 P(f_{i,j}^{(n)}) df_{i,j}^{(n)} = \left[f_{i,j}^{(n)} \right]^{\frac{3}{2}-1} e^{-f_{i,j}^{(n)}}   \,, \label{Tfluct2}
\end{equation}
for each thermofractal $i$ found inside a thermofractal $j$ at level $n-1$, with
\begin{equation}
  f_{i,j}^{(n)}=\frac{F_i^{(n)}}{T_j^{(n)}}\,.
\end{equation}

Suppose now that at the $n$th level the internal energy fluctuations are already small enough to be disregarded and the internal energy is a constant $m_i$. Then according to property 3 of thermofractals, the energy fluctuation of the $j$th thermofractal at the level $n-1$ is proportional to the kinetic energy fluctuation, that is,
\begin{equation}
 P(E_j) \propto \prod_i   \left[f_i^{(n)} \right]^{\frac{3}{2}-1} e^{-f_i^{(n)}+\mu_i} \,,
\end{equation}
where $\mu=m/kT$. But the product of Gamma functions above is itself Gamma fucntion, as described in the Appendix, resulting
\begin{equation}
 P(E_j)dE_j \propto   \left[\frac{F}{kT_j^{(n)}} \right]^{\frac{3N'}{2}-1} exp\left\{-\frac{F}{kT_j^{(n)}}\right\}  exp\left\{-\frac{M}{kT_j^{(n)}}\right\} d\left(\frac{F}{kT_j^{(n)}}\right) \,.
\end{equation}
with $M=\sum m_i$. Since the thermofractals at the $n$th level are being considered as structureless particles, the system at level $n-1$ can be considered as an ideal gas of particles with masses $m_i$. 
The parent thermofractal at level $n-2$ is therefore formed by $N'$ thermofractals, each one considered as an ideal gas of $N'$ particles but at different temperatures $T_j$ and with total energy $M_j$. The probability density to find a set with total internal energy energy $M$ is then
\begin{equation}
 P(M) \propto \int_0^{\infty} \left[\frac{F}{kT_j^{(n)}} \right]^{\frac{3N'}{2}-1} exp\left\{-\frac{F}{kT_j^{(n)}}\right\}  exp\left\{-\frac{M}{kT_j^{(n)}}\right\} d\left(\frac{F}{kT_j^{(n)}}\right) \,. \label{Tfluct3}
\end{equation}
If, at this stage, one still disregards the thermofractal structure, the kinetic energy F can only be interpreted as a parameter, while the system energy  $M$ is the only quantity that keeps some physical meaning, besides the temperature that now fluctuates inside the system. When this step is performed, the equation above is interpreted as a Gamma distribution of the inverse temperature $\beta=1/(kT)$, that is, 
\begin{equation}
 P\left(\frac{1}{T}\right)d\left(\frac{1}{T}\right) \propto  \left[\frac{F}{kT} \right]^{\frac{3N'}{2}-1} exp\left\{-\frac{F}{kT}\right\}  
 d\left(\frac{F}{kT}\right) \,. \label{Tfluct4}
\end{equation}

The distribution of temperatures as described by Eq.~(\ref{Tfluct4}) was already considered in connection to Tsallis distribution in a different context~\cite{BeckTFluct, WilkWlodarczykTFluct, Wilk2018}. On the other hand, the possibility of an equilibrated system with temperature fluctuation is rather controversial~\cite{Kittel1, McFee, Kittel2, Mandelbrot2}. In the present work such fluctuations are well defined in association with the fractal structure of the thermodynamics functions of the system analyzed. Temperature fluctuations arising from a multi scale system was already analyzed in Ref.~\cite{Salazar}.

\section{Callan-Symanzik equation for thermofractals}

Due to the evident similarities between hadron structure and thermofractal structure~\cite{Deppman2012, Megias:2014tha, PedroCardoso, Deppman:2016epj, Deppman2017a} and due to the possible applications of thermofractals or its consequences in Hadron Physics~\cite{Deppman2014,Megias, Megias:2014tha, PedroCardoso}, Astrophysics~\cite{Debora, Megias} and High Energy Physics~\cite{Deppman2012, Sena, Lucas, Lucas2}, it is interesting to approach the thermofractal description in terms close to those used in Quantum Field Theory. This will be done in a future work~\cite{DMMF2}, but it is convenient to advance some aspects here.

The simplest thermofractal diagram corresponds to a vertex with an initial system characterized by $(\vec{\pi}_0,\epsilon_0)$, at an arbitrary level $n$ generating $N'$ systems with $(\vec{\pi}_i,\epsilon_i)$ such that $\vec{\pi}_0=\sum \vec{\pi}_i$ and $\epsilon_0=\sum \epsilon_i$. Such diagram leads to 
\begin{equation}
 P_{i,f} \propto N'^n \prod_{i=1}^{N'} (2 \pi  \epsilon_i)^{-3/2} [\tilde{P}(\varepsilon_i)]^{\nu}\,.
\end{equation}
Here, the passage from one level to the next level represents only an alternative description of the same system. However, one can consider that the initial thermofractal can break into $N'$ pieces, each one being a thermofractal. Let $g$ be a coupling constant that measures such transition from one state to the other, then one can write
\begin{equation}
 \Gamma_{i,j}  \propto N'^n \bar{g}(\epsilon_i) \prod_{i=0}^{N'} (2 \pi  \epsilon_i)^{-3/2}  \,,
\end{equation}
and the term
\begin{equation}
 \bar{g}=  g \prod_{i=0}^{N'} [P(\varepsilon_i)]^{\nu}
\end{equation}
can be considered as an effective coupling constant. $\Gamma_{i,j}$ is, then, understood as a vertex function that is clearly scale free. Vertex functions that are invariant under scale transformation can be described by the Callan-Symanzik equation, that played a fundamental role  in the determination of the asymptotic freedom in Yang-Mills theory. A thermofractal version of such equation was already derived in Ref.~\cite{DeppmanSI}, and it will be derived here in a different way.

The thermofractal temperature $T'=T^{(n)}$ works, as seen above, as a scale parameter that determines the fractal structure level, so one can write the factor $N'^n$ in terms of the level temperature by using Eq.~(\ref{scalingfactor}), i.e.,
\begin{equation}
 N^n \propto T'^{-(1-D)} \,.
\end{equation}
Since $N=N'+2/3$, for the sake of scaling it will be assumed $N \sim N'$, what is a good approximation for $n$ sufficiently high. It results that  the vertex function is
\begin{equation}
 \Gamma_{i,j}  \propto (kT')^{-(1-D)} g \prod_{i=1}^{N'} \left(2 \pi  \frac{E_i}{kT'} \right)^{-3/2}  [P(\varepsilon_i)]^{\nu}\,,
\end{equation}
Notice that when the scale transformation on energy and momentum is performed, so that $\vec{\pi} \rightarrow \lambda \vec{\pi}$ and $\epsilon \rightarrow \lambda \epsilon$, the distribution $P(\varepsilon)$ remains unchanged, since $E/F$ is invariant, therefore it can be left out of the scale invariance analysis of the vertex function studied here. Taking this aspect into account and introducing $M=kT'$ for the sake of simplicity, the scale invariance of the vertex function $\Gamma$ is expressed by
\begin{equation}
 \Gamma(\vec{\pi},E', M') \propto\left(\frac{M'}{M}\right)^{-(1-D)}  \Gamma(\vec{\pi},E, M)\,,
\end{equation}
where it was made use of the scaling property of thermofractals.

From the above expression it is straightforward to conclude that
\begin{equation}
 \begin{split}
  & E_i\frac{\partial \Gamma}{\partial E_i}= - \frac{3}{2} \Gamma \\
  & M \frac{\partial \Gamma}{\partial M} =  \left(\frac{3 N'}{2}-(1-D)\right) \Gamma 
 \end{split}
\end{equation}
and with these results one can write
\begin{equation}
\left[M \frac{\partial}{\partial M}+ \sum_{i=1}^{N'}E_i \frac{\partial}{\partial E_i} + d\right]\Gamma=0\,, \label{CallanSymanzikTsallis}
\end{equation}
where $d=1-D$ is the anomalous dimension for thermofractals, a result equivalent to the one obtained in Ref.~\cite{DeppmanSI}.

The fact that thermofractals satisfy the Callan-Symanzik equation indicates that if it is possible to describe such systems through a field theoretical approach, the Yang-Mills theory is the appropriate framework for it. This results, therefore, sets the grounds for a more fundamental description of thermofractals in terms of gauge field theory, but it will be developed in a future work~\cite{DMMF2}.

\section{Discussion and conclusions}
In the present work the structure of a thermodynamical system presenting fractal structure, recently introduced~\cite{Deppman2016}, is investigated in detail. The fractal structure in thermodynamics has been shown to lead to non extensive statistics in the form of Tsallis statistics, therefore this system can shed some light on relevant aspects of the generalized statistics.

The study presented here evidences the consistency of the proposed
fractal structure of thermodynamical functions that leas to Tsallis
statistics~\cite{Deppman2016}. The diagrammatic representation is a
good auxiliary tool for calculations. In the present investigation the
scaling features of thermofractals are made clear, and it is concluded
that temperature fluctuates from one level of the thermofractal
structure to the other. It is interesting that temperature
fluctuations is pointed as a possible origin of non extensive
statistics~\cite{WilkWlodarczykTFluct}, therefore one can conjecture
that thermofractals will present the same temperature fluctuations
necessary to obtain Tsallis statistics, as given in Eq.(\ref{Trelation}).

One of the main results obtained in the present work 
is given by Eq.~(\ref{Ascale}), showing that the normalizing quantity increases as the system is described by means of structures at deeper levels, $n$. This is a consequence of the fact that the systems at deeper levels contribute less to the energy fluctuation of the system. It follows, on the other hand, from the fact that at deeper levels thermofractals are less massive, and since the energy fluctuation of thermofractals presents self-similarity, energy fluctuation tends to vanish as $n$ increases, as described through Eq.~(\ref{energyscaling}).

Another interesting result is the scale parameter $\lambda_n$, that appears in Eq.~(\ref{scalingfactor}). If $E^{(n)}=\Lambda$ and $E^{(0)}=E$, then it is obtained that
\begin{equation}
 n \log N=(1-D)\log(E/\Lambda)\,. 
\end{equation}
Since $N^n=(N'+3/2)^n \sim M$, for $n$ sufficiently large, with $M$ being the particle multiplicity, it follows that
\begin{equation}
\log M=(1-D) \log(E/\Lambda)\,. 
\end{equation}
Thus the measurement of particle multiplicity gives an easy way to access the fractal dimension $D$.

In addition, it is shown that thermofractals, when the internal structure is not considered, can be interpreted as an ideal gas with inverse temperature that fluctuates according to the Euler's Gamma function. Such temperature distribution was already connected to Tsallis distribution, but here it is obtained as a consequence of the fractal structure of thermodynamics functions.

In Summary, a diagrammatic formulation for calculations with the fractal structure is introduced, what can help in calculations involving several levels of the fractal structure, and some examples are presented. In particular, it is shown the equivalence between tree diagrams and linear diagrams, a result that simplifies the calculations of the relevant quantities. Temperature fluctuations inside the thermofractal is analyzed, reproducing a well-known distribution already connected to Tsallis distribution. The Callan-Symanzik equation for thermofractal structure was obtained, what opens the opportunity to develop a field theoretical approach for thermofractals.

\vspace{0.5cm}
\paragraph*{Acknowledgements}
\thanks{ We are thankfull to Dr. G. Wilk for reading the paper and giving interest suggestions.
A.D., D.P.M. and T.F. are partially supported by Conselho Nacional de
Desenvolvimento Cient\'{\i}fico e Tecnol\'{o}gico (CNPq-Brazil) and  
by  Project INCT-FNA Proc. No. 464898/2014-5. The work of EM is supported by the Spanish MINEICO under Grants FPA2015-64041-C2-1-P and FIS2017-85053-C2-1-P, by the Basque Government under Grant IT979-16, by the Junta de Andaluc\'{\i}a under Grant FQM-225, and by the Spanish Consolider Ingenio 2010 Programme CPAN (CSD2007-00042). The research of EM is also supported by the Universidad del Pa\'{\i}s Vasco UPV/EHU, Bilbao, Spain, as a Visiting Professor, and by the Ram\'on y Cajal Program of the Spanish MINEICO.
}

\vspace{1cm}

\appendix
\section{Appendix: Useful formulae}

The energy distribution of an ideal gas is given by
\begin{equation}
 \int_{0}^{\infty}P(F) dF=C \int_{-\infty}^{\infty} d^{3N'}p_i \exp\left[-\beta \sum_{i=1}^{3N'} \frac{p_i^2}{2m} \right]\,, \label{uf1}
\end{equation}
where $\beta=1/(kT)$ and $C$ is the normalization constant. If the momenta of the different particles are independent, then
\begin{equation}
 \int_{0}^{\infty} P(F) dF=C \left\{\int_{-\infty}^{\infty} dp_i \exp\left[-\beta \frac{p_i^2}{2m} \right]\right\}^{3N'}\,.
\end{equation}
Since 
\begin{equation}
 \int_{-\infty}^{\infty} dp_i \exp\left[-\beta \frac{p_i^2}{2m} \right]= (2 \pi m kT)^{1/2} \,,
\end{equation}
the normalization constant must be chosen as
\begin{equation}
 C=(2 \pi m kT)^{-3N'/2}\,.
\end{equation}
Notice that the integration in Eq.~(\ref{uf1}) can be performed in terms of the total momentum $p^2=\sum_{i=1}^{3N'} p_i^2$, by considering a hypersphere of dimension $n=3N'$. Then
\begin{equation}
 \int_{0}^{\infty} P(F) dF=C \int_{0}^{\infty} dp \, S_n p^{n-1} \exp\left[-\beta \frac{p^2}{2m} \right]\,, \label{uf2}
\end{equation}
where
\begin{equation}
 S_n=\frac{2 \pi^{n/2}}{\Gamma(n/2)}
\end{equation}
is a surface factor for the $n$-dimensional hypersphere, with $\Gamma$ being the Euler Gamma Function.
But $F=p^2/(2m)$, then
\begin{equation}
 dF=\frac{p}{m}dp \,,
\end{equation}
hence
\begin{equation}
 \frac{dF}{2F}=\frac{dp}{p}\,. \label{uf3}
\end{equation}
Therefore from Eq.~(\ref{uf2}) one has
\begin{equation}
 \int_{0}^{\infty} P(F) dF=C S_n \frac{(2m)^{n/2}}{2} \int_{0}^{\infty} \frac{dF}{F} F^{n/2}  \exp(-\beta F) \,.
\end{equation}
Substituting the relations for $S_n$, $n$ and for $C$, it results
\begin{equation}
 \int_0^{\infty} P(F) dF= \int_0^{\infty} \frac{1}{\Gamma(3N'/2)}  \frac{dF}{kT} \left(\frac{F}{kT}\right)^{(3N'/2)-1} \exp(-\beta F) \,. \label{uf4}
\end{equation}

\end{document}